# Motion correction for brain MRI using deep learning and a novel hybrid loss function

Lei Zhang[1], Xiaoke Wang[1], Michael Rawson[2], Radu Balan[3], Edward H. Herskovits[1], Elias Melhem[1], Linda Chang[1], Ze Wang[1], and Thomas Ernst[1]

[1]Department of Diagnostic Radiology and Nuclear Medicine, University of Maryland School of Medicine, Baltimore, MD, USA

[2]Pacific Northwest National Laboratory

[3]Department of Mathematics and Center for Scientific Computation and Mathematical Modeling, University of Maryland, College Park, MD, USA,

**Corresponding Authors**

Thomas Ernst, Ph.D.

Department of Diagnostic Radiology and Nuclear Medicine

University of Maryland School of Medicine

670 W. Baltimore Street, HSF-III, Room 1130,

Baltimore, MD 21202

Email: ternst@som.umaryland.edu

Ze Wang, Ph.D.

Department of Diagnostic Radiology and Nuclear Medicine

University of Maryland School of Medicine

670 W. Baltimore Street, HSF-III, Room 1130,

Baltimore, MD 21202

Email: ze.wang@som.umaryland.edu

Figure and Tables: 9 Figures and 1 Table (1 supplementary table)

**Key words:** MRI, motion correction, deep learning, brain

# Abstract

## Purpose

To develop and evaluate a deep learning-based method (MC-Net) to suppress motion artifacts in brain magnetic resonance imaging (MRI).

## Methods

MC-Net was derived from a UNet combined with a two-stage multi-loss function. T1-weighted axial brain images contaminated with synthetic motions were used to train the network. Evaluation used simulated T1 and T2-weighted axial, coronal, and sagittal images unseen during training, as well as T1-weighted images with motion artifacts from real scans. Performance indices included the peak signal to noise ratio (PSNR), structural similarity index measure (SSIM), and visual reading scores. Two clinical readers scored the images.

## Results

The MC-Net outperformed other methods implemented in terms of PSNR and SSIM on the T1 axial test set. The MC-Net significantly improved the quality of all T1-weighted images (for all directions and for simulated as well as real motion artifacts), both on quantitative measures and visual scores. However, the MC-Net performed poorly on images of untrained contrast (T2-weighted).

## Conclusion

The proposed two-stage multi-loss MC-Net can effectively suppress motion artifacts in brain MRI without compromising image context. Given the efficiency of the MC-Net (single image processing time ~40ms), it can potentially be used in real clinical settings. To facilitate further research, the code and trained model are available at https://github.com/MRIMoCo/DL_Motion_Correction.

## Introduction

Magnetic resonance imaging (MRI) is a widely used medical imaging modality due to its ability of visualizing both the anatomy and function of tissues and organs as well as pathologic processes[1]. MRI provides high spatial resolution and diverse contrasts, making it superior to many other imaging modalities for detecting and characterizing soft tissue (e.g., brain, abdominal organs, and blood vessels) and pathologies.

Because of the sequential spatial encoding steps used to spatially encode the imaging object, MRI is relatively slow and can take up to several minutes for a typical 3D volume scan. The prolonged image acquisition process makes MRI sensitive to motion[1,2]. Unfortunately, motion in human subjects is inevitable and can be caused by involuntary physiological movements, such as respiration and cardiac motion, and unintended patient movements. Motion-induced image artifacts can drastically deteriorate image quality and reduce diagnostic accuracy[2]. For example, Andre et al reported that almost 60% of 192 clinical brain MRI scans were contaminated with motion artifacts[3]. Among these, 28% were marginally diagnostic to non-diagnostic and should be repeated. Because of the motion-induced image artifacts, the annual loss of revenues per MR scanner can be over $100,000 for brain studies alone[3].

A range of prospective correction strategies have been developed for motion artifacts [4–7 8], but they commonly have limitations such as scanner platform accessibility, applicability to specific MR imaging sequences, and limitations in correcting different types of motion artifacts (e.g., in-plane versus through-plane movements). Therefore, retrospective motion-correction by means of post-processing provides a good complement. One promising approach involves deep learning (DL)[1,2,9–16,17,18], using deep convolutional neural networks (DCNN) or other network architectures with supervised learning. Given sufficient training pairs (inputs and reference images), DCNNs can be trained to learn the transformation from the input (motion-corrupted image) to the reference (motion-free image). Trained DCNNs have been used successfully to solve many challenging and clinically important problems, e.g., arterial spin labeling perfusion MRI denoising [13,19], image segmentation [20,21], and image registration [22,23].

DCNNs appear to be well-suited for retrospective correction of motion artifacts since there are no obvious conventional algorithms to solve the problem and yet expert readers can "read through" the artifacts to some degree. Recent studies demonstrate that DCNNs can be used to attenuate motion artifacts in brain MRI scans using a data-driven approach without prior knowledge. For instance, variation auto encoders (VAE) and generative adversarial networks (GAN) were implemented for retrospective correction of rigid and non-rigid motion artifacts from motion-affected MR scans[1]. GANs were also used for motion correction in multi-shot MRI [12]. A conditional GAN improved the image quality of predicted motion-corrected images compared to motion-corrupted images[24]. Finally, an encoder-decoder network was able to suppress motion artifacts with motion simulation augmentation[2].

The purpose of this study was to implement and comprehensively evaluate a new deep neural network architecture and loss function for motion correction. The methodology and scope of this study are different from previous studies. The novel and unique contributions of this paper include: First, a new loss function was proposed, which contained an L1 component for penalizing overall image artifacts and a total variation component to penalize the loss of image details such as boundaries. Accordingly, a two-stage training strategy was implemented to first minimize the overall motion-artifacts and then consider both the residual motion-induced artifacts and the loss of image details such as boundaries. Second, the generalizability of the trained model was assessed using images with different contrast than that of the training data. Third, to ensure rigor and demonstrate clinical utility, substantial evaluations were made using different levels of synthetic motions and in-vivo data through both objective performance indices and subjective reading by experienced clinicians. Motion-free images were also used to assess potential over-corrections by the trained DL networks. Finally, to allow other researchers to reproduce our work or use the methods to process their own data, we have released the code and sample data at https://github.com/MRIMoCo/DL_Motion_Correction.

## Methods

### 2.1 MC-Net

The proposed deep learning-based method (MC-Net) takes a motion-corrupted image as input and outputs a motion-corrected image. The method implements a modified UNet (Figure 1) as its neural-network structure. The MC-Net was trained with a novel two-stage training strategy using a hybrid loss function L that combines a L1-loss and Total Variation loss (TV) [25]:

$$L = \text{alpha} * L1 + \text{beta} * TV \quad (1)$$

$$L1 = \sum_{i,j} |I(i,j) - I_0(i,j)| \quad (2)$$

$$\text{TV} = \sum_{i,j} \left( \left(I(i+1, j) - I(i,j)\right)^2 + \left(I(i, j+1) - I(i,j)\right)^2 \right)^{1.25} \quad (3)$$

where $I$ and $I_0$ are a corrupted image and a motion free image, and $i$ and $j$ are row/column indices. During the first training stage, we used the L1-loss only [(alpha, beta) = (1, 0)] to suppress overall motion-induced artifacts. The pre-trained stage 1 model is then fine-tuned in stage 2 by turning on the TV-loss component [(alpha, beta) = (1, 1)]; this penalizes boundary artifacts in addition to the overall artifacts.

## 2.2 Motion Corrupted Images

The pipeline that generates motion-corrupted k-space data is shown in Figure 2. The project used images with simulated motion artifacts, based on deidentified brain MRI scans from 52 human subjects (50 male, 2 female, age 48.6 ± 9.1 years) previously enrolled in research studies. All data were acquired on a 3T scanner (TIM Trio, Siemens Healthcare, Erlangen, Germany). The ability of the MC-Net to correct real (non-simulated) images with motion artifacts was assessed using motion-corrupted scans from five additional subjects (2 male, 3 female, age 19±4.9).

The source images were 3D sagittal magnetization-prepared rapid gradient-echo (MP-RAGE) scans and 2D axial fluid-attenuated inversion-recovery (FLAIR) scans from 52 subjects. MP-RAGE images were collected with the following parameters: TR=2.2s; TE=4.47ms; TI=1s; resolution=1mm isotropic; matrix size=256×256×160, and FLAIR images were collected with the following parameters: TR=9.1s; TE=84ms; echo train length=11; matrix size=256×204; in-plane resolution = 1mm$^2$; slice thickness=3mm; slice spacing=3mm; TI=2.5s. All source images were assessed visually to ensure they did not contain motion artifacts.

Forty-two axial in-plane motion trajectories of 256 temporal samples each were synthesized from in-vivo head movements measured with the prospective acquisition correction (PACE) algorithm[27] during BOLD functional MRI (fMRI) scans. The source motion trajectories had translations <2mm and rotations <2°, and were subsequently multiplied by eight and reduced from 6 degrees of freedom to in-plane motions (3 degrees of freedom). The offsets of the trajectories were normalized such that the motions in all axes are zero at the center of the trajectory. The severity of each motion trajectory applied is indicated by the motion standard deviation across time for all three in-plane degrees of freedom (L2-norm of in-plane translations in mm and in-plane rotation in degree).

To simulate motion artifacts, the 3D MP-RAGE images were zero-padded to 256×256×256, and sagittal as well as re-sliced axial and coronal images were extracted. The resulting 2D images were then Fourier-transformed to create k-space data. Rigid motion corrupts k-space data by changing the sampled k-space trajectory and accumulating additional phases in each k-space sample. For each of the 42 trajectories, the associated k-space trajectory was calculated using the augmented homogeneous transform[8]. Next, the k-space data of the non-corrupted 2D images were re-sampled on the corrupted k-space trajectory with non-uniform FFT (NuFFT) [28], and additional phase ramps due to translations were added to obtain corrupted images after inverse Fourier transformation.

Four different datasets were used in this work. Datasets 1, 2, and 3 were generated with simulated motion, whereas dataset 4 contains images with real motion artifacts from patient movement.

**Dataset 1** was used to train the neural network. Ten axial MP-RAGE slices spaced 5mm apart were extracted from each "clean" (original) and motion-corrupted MP-RAGE volume. These images were divided into a training set (35 subjects; 25 motion trajectories; 13,700 slices), validation set (5 subjects; 7 motion trajectories; 1,950 slices), and test set (12 subjects; 7 motion trajectories; 4,680 slices). The subjects and motion trajectories in these sets did not overlap.

**Dataset 2** was used to test the generalizability of MC-Net to unseen anatomical structures. It comprised motion-corrupted MP-RAGE images in sagittal (140 total slices) and coronal (140 total slices) views of five (of 12) subjects from the test group.

**Dataset 3** was used to test how well the MC-Net adapts to a different image contrast without additional training. It consists of motion-corrupted axial FLAIR images from five (of 12) subjects in the test group.

**Dataset 4** was used for testing the MC-Net using data with real, rather than simulated, motion. It includes a total of 15 T1-weighted images from five subjects, three of whom moved during the scan and were not included in the test data set.

*2.5 Quantitative evaluation metrics*

The performance of the various methods implemented was quantified using two measures. First, using the "clean" image as reference, the structural similarity index measure (SSIM)[29] was calculated:

$$SSIM(x,y) = \frac{(2\mu_x\mu_y+C_1)(2\sigma_{xy}+C_2)}{(\mu_x^2+\mu_y^2+C_1)(\sigma_x^2+\sigma_y^2+C_2)} \tag{4}$$

where x and y are signals from input and reference, and $\mu_x$, $\mu_y$, $\sigma_x$, $\sigma_y$, and $\sigma_{xy}$ denote mean of input, mean of reference, variance of input, variance of reference, and covariance. $C_1$ and $C_2$ are two constants to keep the denominator from becoming zero.

The second performance measure was the peak signal to noise ratio (PSNR). The PSNR is the ratio of maximum possible power (MAX) of a signal and the power of distorting noise that affects the quality of its representation (mean square error between the denoising result from a DL model and the clean reference, MSE):

$$\text{PSNR} = 10\log_{10}\left(\frac{\text{MAX}^2}{\text{MSE}}\right) \tag{5}$$

When the numerical difference between a predicted image and the reference approaches 0, the PSNR approaches infinity and the SSIM approaches 1.

*2.6 Visual Reading Scores*

The artifacts in the clean (reference) images, motion-corrupted images, and MC-Net predicted images were assessed visually by two experienced imaging specialists who were blinded for the correction method. The scores for image artifacts ranged from 0 to 3 (0 = "no"; 1 = "mild"; 2 = "moderate"; and 3 = "severe" motion artifacts).

For simulated motion experiments, a mid-axial slice of five subjects from the test group were selected. Ten distinct motion trajectories with a wide range of summed standard deviations were selected from the test group. The clean image, motion-corrupted image, and MC-Net prediction of each possible pair of a subject and a motion trajectory were reviewed by the imaging specialists in a randomized order.

For real (non-simulated) motion experiments, the motion-corrupted images and the MC-Net predictions of the five subjects in dataset 4 were reviewed. Three slices were reviewed per subject, and motion artifacts were rated using the scale described above.

The artifact scores of motion-corrupted images and MC-Net predictions were compared with the one-sided Wilcoxon signed-rank test, after averaging the reader scores. For simulated motions, artifact scores were compared for three ranges of summed standard deviation: [0,5](mm/°), (6,10](mm/°), and (10,15](mm/°).

*2.7 Implementation Details*

All neural networks in this work were implemented in Keras [30], a deep learning programming platform built upon TensorFlow [31]. Our code supports Nifti image format. A workstation with an Intel® Core™ i7-9700K CPU (3.60GHZ, 64 GB RAM) and two Nvidia GeForce Titan 2080 GPU was used.

All models were trained in 200 epochs with a batch size of 8. We used the Adam optimization method[32] with initial learning rate of 0.0001 and periodic decay = 0.96 after every epoch [2]. An early stopping technique[33] was used during all the DL models training to avoid overfitting. Experiments were performed with two slightly different network structures (Fig. 1). The first experiment involved a conventional UNet structure ("U") and was used for the main results presented here (including visual evaluation by radiologists). The other experiment involved an additional connection from input to output ("U+O") to preserve more information from the input images[2] and was used for comparison with the "U" experiment. For comparison with the final MC-Net with the two-stage multi-loss function, single-stage models trained with the L1-loss [2] and with L1 + TV loss were also implemented ("L1" and "L1 + TV" models).

Finally, since the mapping provided by DL algorithms is intrinsically non-linear, we performed an additional test to assess the preservation of information by feeding motion-free images into the various models and calculate the difference between the predicted results and the motion-free source images (see supplementary).

## Results

To test the processing speed of the final network, 1000 images were submitted to MC-Net. This demonstrated that our final algorithm could process one image in 40ms on average.

*3.1 Quantitative improvements for motion-corrupted images*

Table 1 shows the SSIM (mean ± std) and PSNR (mean ± std) values of motion-corrected images compared to uncorrected images in the test set. The two-stage solution had the best performance among the 3 algorithms implemented (L1; L1 + TV; two-stage), although commonly only by a relatively small margin.

In comparison to the previously published UNet-like structure with optional input-to-output concatenation ("U+O")[2], the proposed UNet solution showed improved performance across the various methods and measures. This difference was especially pronounced for the SSIM, which improved only marginally from 0.773 (corrupted images) to 0.816 (corrected images) for the U+O method with two stages, but to 0.919 (corrected images) for the two-state algorithm without optional concatenation (U).

Figure 3 shows the relation between SSIM and the motion magnitude (standard deviation across 256 time points) for corrupted images and images corrected with the MC-Net algorithm (two-stage multi-loss function). Without correction of motion artifacts, the SSIM decreases with the motion magnitude (from >0.95 for small motions below 1.5mm/°) to approximately 0.75 for severe motion (slope of linear regression curve = -0.028). The MC-Net consistently improved image quality but was especially effective for large movements (slope = - 0.014), such that the effect of motion on image quality decreased 2-fold (slope of regression line).

### 3.2 Effects on artifact-free images

When artifact-free images were processed with the various networks, all algorithms generated high quality images, with SSIMs consistently above 0.95 and almost 0.97 for the two-stage multi-loss function UNet. However, the network with optional input-to-output concatenation achieved almost perfect agreement between the input and output images (SSIM>0.99). One example is shown in Figure 4, where the SSIM of a minimally corrupted image improves from 0.95 to 0.97 after processing with the MC-Net.

Overall, the UNet with two-stage multi-loss function (MC-Net) yielded the best performance for correction of corrupted images while retaining the most information for "clean" images.

### 3.3 Visual reading

Fig. 5 summarizes the readers' visual assessments of reference images, motion-corrupted images, and the MC-Net predictions against the motion magnitude for 10 selected motions each for 5 subjects (from the "test" data set with simulated movements). The "clean" reference images were consistently scored as showing no or only minor artifacts (red lines). Conversely, scores of corrupted images consistently worsened with the degree of motion, such that most images with >7.5mm/° motion (standard deviation across scan) were rated as having "moderate" or "severe" artifacts (blue lines). The visual scores of images processed with MC-Net improved significantly for motions between [0,5]mm/° (p=5·$10^{-4}$ on Wilcoxon signed-rank test), [5,10]mm/° (p=6·$10^{-5}$), as well as between [10,15]mm/° (p=1.9·$10^{-3}$). In fact, the average quality of corrected images was in the "minor artifacts" level even for the most severe range of motion (>7.5mm/° motion; green versus blue lines).

Representative results of the L1, L1+TV, and MC-Net algorithms are shown in Figure 6. All DL methods improved image quality, but the MC-Net was slightly more effective than the other methods. For instance, in an image with moderate motion artifacts (Fig.6A, top panel), the MC-Net improved the SSIM from 0.72 to 0.92 and the PSNR from 22.9 to 27.9.

An image with more severe motion artifacts is shown in Figure 6B, bottom panel. The corrupted image had an SSIM of 0.61 and PSNR of 22.31 due to the severe motion applied (10-15mm/° sudden jump in middle of scan). After motion correction, the SSIM and PSNR improved to 0.88 and 27.06, respectively, for MC-Net.

Of note, while motion artifacts were suppressed substantially after processing images with originally moderate to severe artifacts, the corrected (output) images appeared to be processed with a low-pass filter. This contrasts with input images with little or no artifacts, where the low-pass effect was absent from output images (e.g., Figure 4).

*3.3 Cross-dataset Generalization*

To assess cross-dataset generalization, we first applied the MC-Net algorithm (which was trained using axial images only) to sagittal and coronal T1-weighted slices with simulated motion artifacts. Representative results from two different scans are shown in Figure 7. The images from the first subject (Fig. 7, top row; 3.95mm/°) show mild to moderate motion artifacts. The initial SSIM values are similar for the two views, both before (0.83 and 0.84) and after elimination of artifacts with the MC-Net (0.87-0.9). Images from the second subject (Fig. 7, bottom row; 5.99mm/°) show moderate to severe motion artifacts, and SSIM values improve from 0.49 to 0.73 (sagittal) and 0.68 to 0.84 (coronal). Overall, these improvements in SSIM scores after processing with the MC-Net are consistent with those observed for axial scans (regression lines in Figure 3).

To test the ability of the MC-Net algorithm to correct motion artifacts in T2-weighted images, we randomly selected and corrupted 264 slices from the FLAIR dataset. The SSIM and PSNR of the corrupted images were (0.69 ± 0.11) and (26.15 ± 2.72) (mean ± std), compared to (0.67 ± 0.04) and (25.25 ± 1.54) after correcting. Thus, processing with MC-Net overall decreased the image quality.

This finding is confirmed visually on two axial slices from two different subjects (Fig. 8) using simulated motion artifacts. In the first set of images (Fig. 8A) with relative minor motion (3.77 mm/°), the SSIM and PSNR values are poorer after processing with the MC-Net and slight banding artifacts appear. Conversely, when the MC-Net was applied to a FLAIR image with more severe motion artifacts (Fig. 8B, 14.85 mm/°), the SSIM and PSNR values were improved. However, despite these apparent improvements, the zoomed region of the "corrected" image provides a poor representation of the original scan; in fact, new (false) anatomical "features" appear.

*3.4 Images with real (non-simulated) motion*

As a final validation step, we applied the MC-Net algorithm to human brain images with various degrees of motion artifacts (not simulated). Some representative results are shown in Figure 9, with the apparent degree of motion-corruption increasing from (A) to (C). As before, the image with minor motion artifacts (9A) was well preserved by the MC-Net, whereas the two scans with moderate and severe artifacts (9B) and (9C) showed substantial improvements in quality after processing with the MC-Net. While obviously no reference images are available to quantify improvements in image quality, visual image scores by our blinded readers improved significantly between original and corrected scans (p=0.04).

## Discussion

The goal of our work was to develop a DL-based automated method to eliminate or attenuate motion-artifacts from brain MRI scans, the MC-Net. A novel loss function and novel training strategy were proposed. The proposed MC-Net network was trained and tested using data with a wide range of real movements. A series of comprehensive evaluations were performed to assess the risk of over-correction and generalizability. The former was conducted by inputting the motion-free images and assess how well the quality of images was preserved after being processed by MC-Net. The latter was performed through testing MC-Net on a different dataset with totally different image orientation and image contrasts.

In agreement with previous work[2,11,12], the MC-Net was able to correct motion artifacts in brain scans. We based our work on a customized UNet-like auto-encoder architecture [34] that is widely used in medical image analysis. The network architecture used is complex enough to perform the motion correction task, but also simple enough to benefit from the hybrid loss function. The proposed hybrid loss and two-stage training was slightly better in improving the final image quality than approaches with single stage training, e.g., the L1 loss used in a prior study [2], and was especially beneficial for correcting severe motion artifacts. Due to the short processing time per image (40ms), implementation using the open-source framework Keras [30] with GPU acceleration should allow real-time motion correction.

*4.1 Advantages of the Two-stage Training and Multiple-loss function*

Two-stage training can distill different information at each stage and benefit applications such as reconstruction of high resolution arterial spin labeling MRI [35] and image denoising [36]. We observed that the first L1 stage can recreate the overall anatomical structure, but some fine details were missing when large movements were present. Inspired by the style transfer approach [25,37], we added the total variation term to remove residual artifacts and introduce smoothness across pixels. As a result, the MC-Net is slightly better than the multi-loss method with single-stage training, perhaps since the model trained in one stage only may fall into a local minimum, which the second stage training step might have avoided.

*4.2 Comparison of Different DL Architectures*

We tested two different architectures, i.e., UNet-like (U) and Unet-like plus an additional input-to-output connection (U+O) as suggested in prior work [2]. Interestingly, the simpler U-model yielded better SSIM and PSNR values than the U+O model on motion-corrupted images. Therefore, the U-architecture with the two-stage multiple-loss method had the overall best performance for both motion-corrupted and motion-free images.

All three U+O models [2] performed perfectly in terms of SSIM for motion-free images (SSIM essentially 1.0), independent of which loss function was used during training. This is likely a result of the additional input-to-output connection in the U+O model, which allowed passing of more information from the original image to the output compared with the U model. This additional information made the results from U+O more like the motion-free image.

However, amongst the U-models, the model trained with the two-stage multiple-loss method still had the best PSNR for motion-free images, which demonstrates that the two-stage multiple-loss method might be applied widely to different DL architectures.

*4.3 Performance on Test Set*

MC-Net achieved promising results on corrupted axial test datasets with T1-contrast, both in terms of SSIM and PSNR. In addition, the radiologic scores from two experienced imaging specialists demonstrate that MC-Net can attenuate motion artifacts and improve image quality.

*4.4 Cross-dataset generalization*

Our DL model was trained with a single orientation (axial) and contrast (T1-weighting using MP-RAGE). Therefore, we assessed the generalizability of the method using datasets with untrained orientations and untrained contrast. In terms of new orientations (sagittal and coronal) without change in contrast (T1), quantitative improvements in image quality measures after processing with the MC-Net were similar to those for the trained orientation (axial) for a given level of motion artifacts. Importantly, the MC-Net also improved the quality of axial T1-weighted images that were corrupted during acquisition. However, we did not evaluate the performance for T1-weighted images acquired with different pulse sequences, such as FLASH or spin-echoes.

Conversely, the MC-Net provided little to no benefit in terms of correcting motion artifacts in axial T2 view images. Therefore, the image contrast appeared to have more relevance in representing imaging artifacts than anatomical structure. Therefore, motion correctio may be considered a low-level vision problem in which a low-level clue (contrast) is more important than a high-level clue (anatomical structure).

Compared with previous studies using DL to correct motion-induced artifacts, our paper contains a comprehensive and in-depth assessment of the performance of the MC-Net. In addition to quantitative measures such as SSIM and PSNR, the severity of motion artifacts was assessed by imaging specialists and analyzed statistically. The motion trajectories in the simulated motion experiment span a wide range of severity and were synthesized from real recorded motion trajectories in exams. The MC-Net was tested for cross-dataset generalization with respect to the anatomical structures, which changes with image orientation, and to the image contrast.

*4.5 Limitations*

While we attempted to perform a comprehensive evaluation of the proposed MC-Net, our study has also several limitations. First, the range of imaging contrasts evaluated was focused on T1-weighted, and more specifically MP-RAGE, and a few T2-weighted scans (FLAIR). Therefore, we do not know the performance of our method for other contrasts, such as susceptibility weighting or conventional T2-weighted contrast. However, given the poor performance of the MC-Net in processing the FLAIR images, it is likely that the current algorithm would not be beneficial for other contrasts either. Likewise, the images in our training data set were relatively normal in terms of anatomy, and the performance of our method still must be assessed in scans with brain lesions or morphometric abnormalities.

Second, we did not include sagittal and coronal data in the training set and evaluate the performance, since we focus on a scenario where only one source of data is available and sufficient to train the model, i.e., single domain generalization[38]. These properties might be relevant in the clinical setting.

The motion trajectories used in this work were based on real head motions measured with PACE-fMRI, and are more realistic compared with options such as sinusoids and other regular patterns or random motion trajectories. However, the time scale of the motion trajectories was not resampled to match that of the MP-RAGE and FLAIR acquisitions. Consequently, movements in the simulations were in effect slowed down or accelerated compared with the PACE measurements. Further, the amplitude of movements was magnified to create more challenging artifacts for the MC-Net. Overall, some authenticity of the motion trajectories was sacrificed for higher temporal resolution and the higher motion amplitudes.

It is worth to note that the model has not been trained and tested for extremely large motions. Adding extreme cases might done by fine-tuning the model with new training samples and intermediate samples, possibly to be generated using a Projection-On-Convex-Set based cycle Generative Adversarial Network [39].

## Conclusions

We developed a simulation framework and MC-Net for correction of motion artifacts in brain MRI. The MC-Net performed well on unseen data and images of different orientations with the same contrast (T1-weighted), but scans with other contrast will require additional optimizations. The high reader scores and evaluation metrics of corrected images demonstrate the viability of the MC-Net for correcting motion artifacts. Since the method is data-driven, independent of data acquisition or reconstruction and fast to perform, it may ultimately be suitable for routine clinical practice.

## Acknowledgements

This work was supported by NIH grant 1R01 DA021146 (BRP), R01AG060054, R01AG070227, R01EB031080-01A1, and P41EB029460-01A1. M.R. and R.B. were supported in part by NSF grant DMS-2108900 and Simons Foundation Fellowship grant 818333. We also thank Dr. Pan Su from Siemens Healthineers for his valuable suggestions.

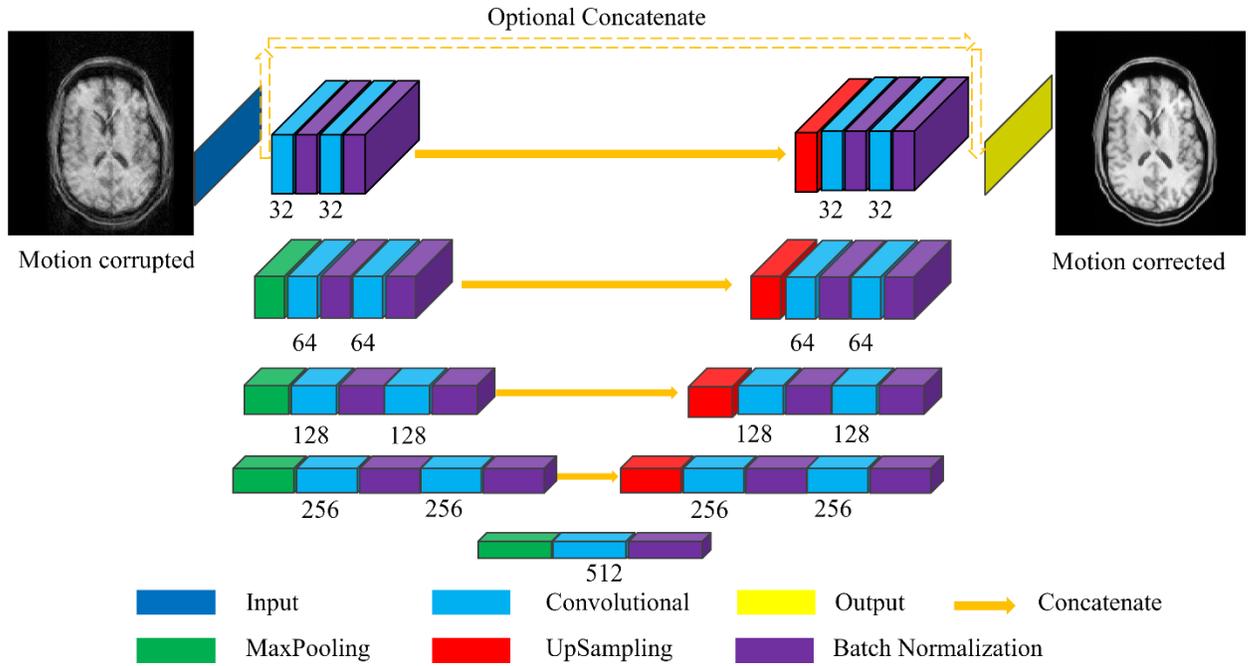

Figure. 1: Architecture of the MC-Net, which was derived from UNet. The filter number in each convolutional layer of the customized UNet is half of the original UNet [26].

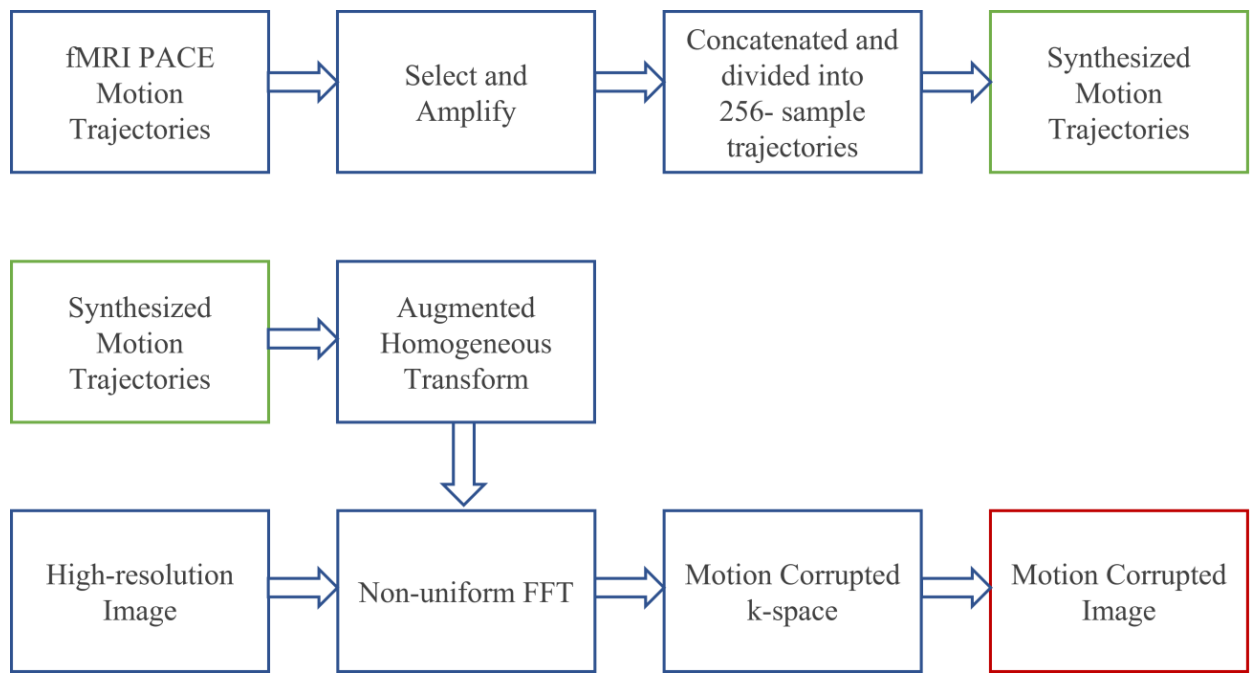

Figure 2. The pipeline to generate motion-corrupted k-space data.

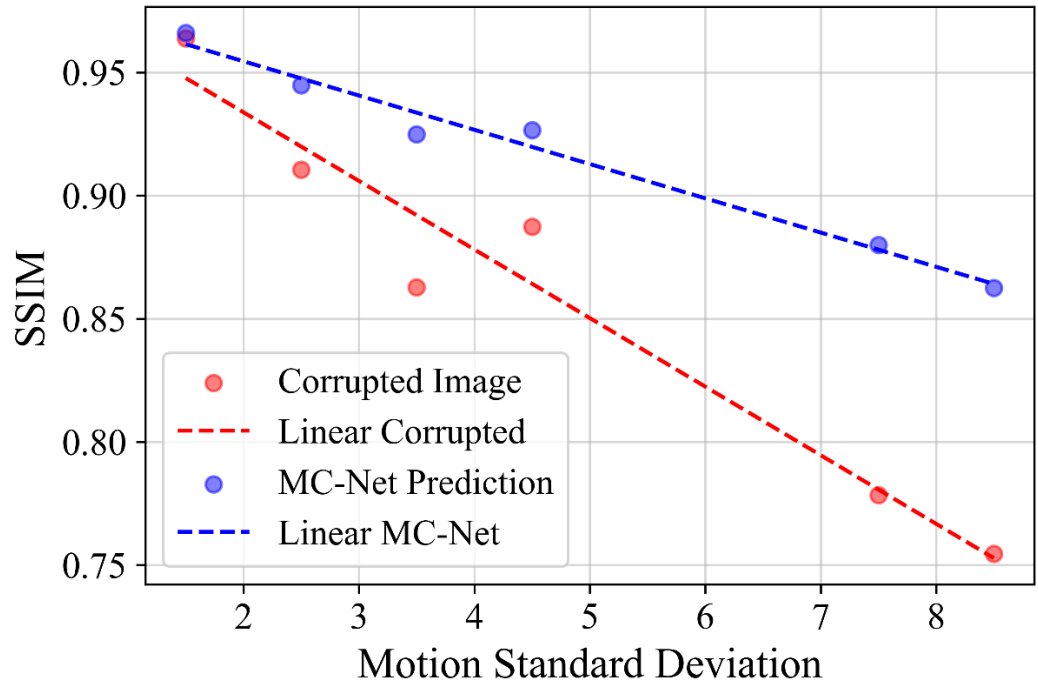

Figure 3: SSIM values of images corrected with MC-Net (blue dots) relative to those of uncorrupted images (red dots) are plotted against the magnitude of motion simulated (standard deviation of motion across 256 time points, in mm or mm/°). The red line ($Y = 0.99 - 0.028X$) and blue line ($Y = 0.98 - 0.014X$) show linear regression of images without and with motion correction, respectively.

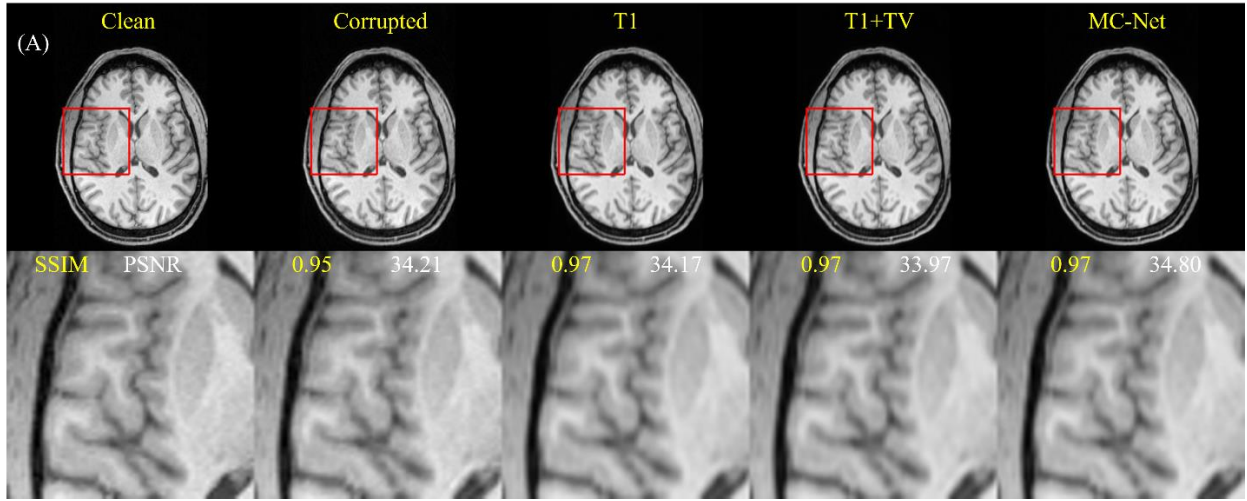

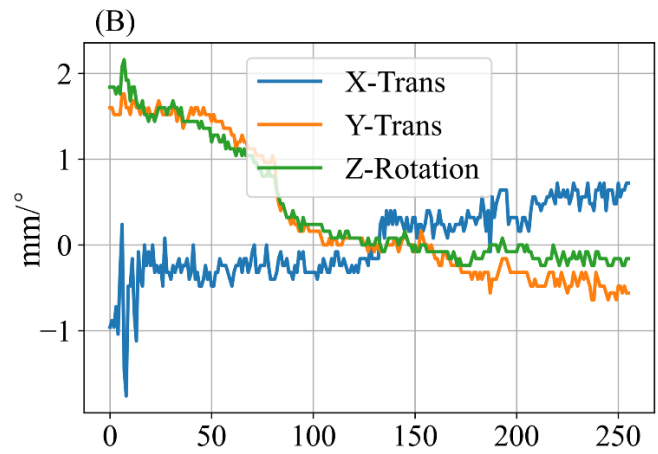

**Figure 4:** Examples of motion artifact removal from images in the test data set using various algorithms. In Figure 4A (1.90 mm/°), the first row contains the clean reference image, corrupted image, and motion correction results using the L1, L1+TV, and MC-Net algorithms. The second row shows an enlarged image of the red rectangle. The SSIM and PSNR values for each corrected image (relative to the "clean" image) are also shown (bottom row). Figure 4B shows the motion trajectory for Fig 4A, where the horizontal axis labels refer to y-position in k-space.

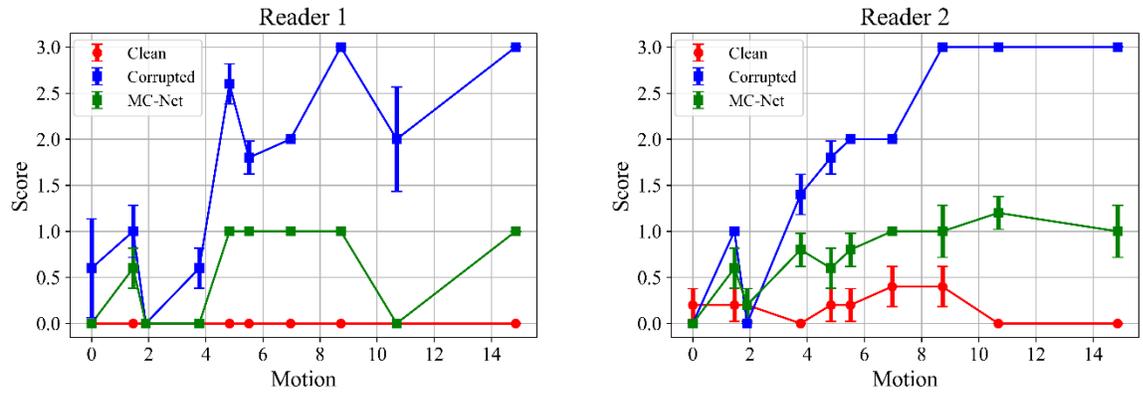

**Figure 5:** Average motion artifact scores from two blinded clinical readers (left and right graph) for "clean" reference images (red lines), motion-corrupted images (blue lines), and the MC-Net predictions (green lines). The x-axis represents the standard deviation of motion (in 5mm/°), and the y-axis shows average reading scores. Error bars represent standard errors of the mean.

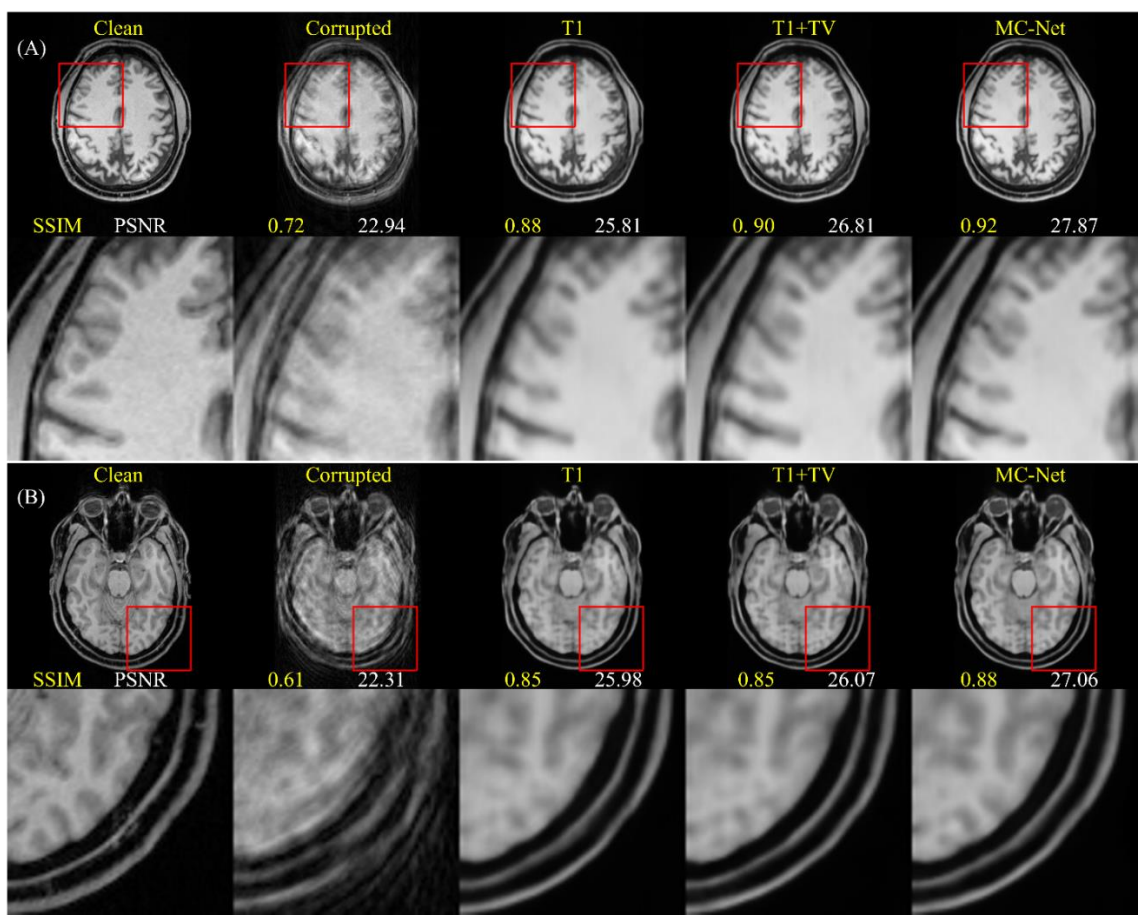

**Figure 6:** Examples of motion artifact removal from images in the test data set using various algorithms. In Figure 6A (4.67 mm/°) and 6B, the first row of each subfigure contains the clean reference image, corrupted image, and motion correction results using the L1, L1+TV, and MC-Net algorithms. The second row of each subfigure zooms in on the red rectangle. The SSIM and PSNR values for each corrected image (relative to the "clean" image) are also shown (bottom row). Figures 6C and 6D show the motion trajectories for Fig 6A and Fig 6B, where the horizontal axis labels refer to y-position in k-space.

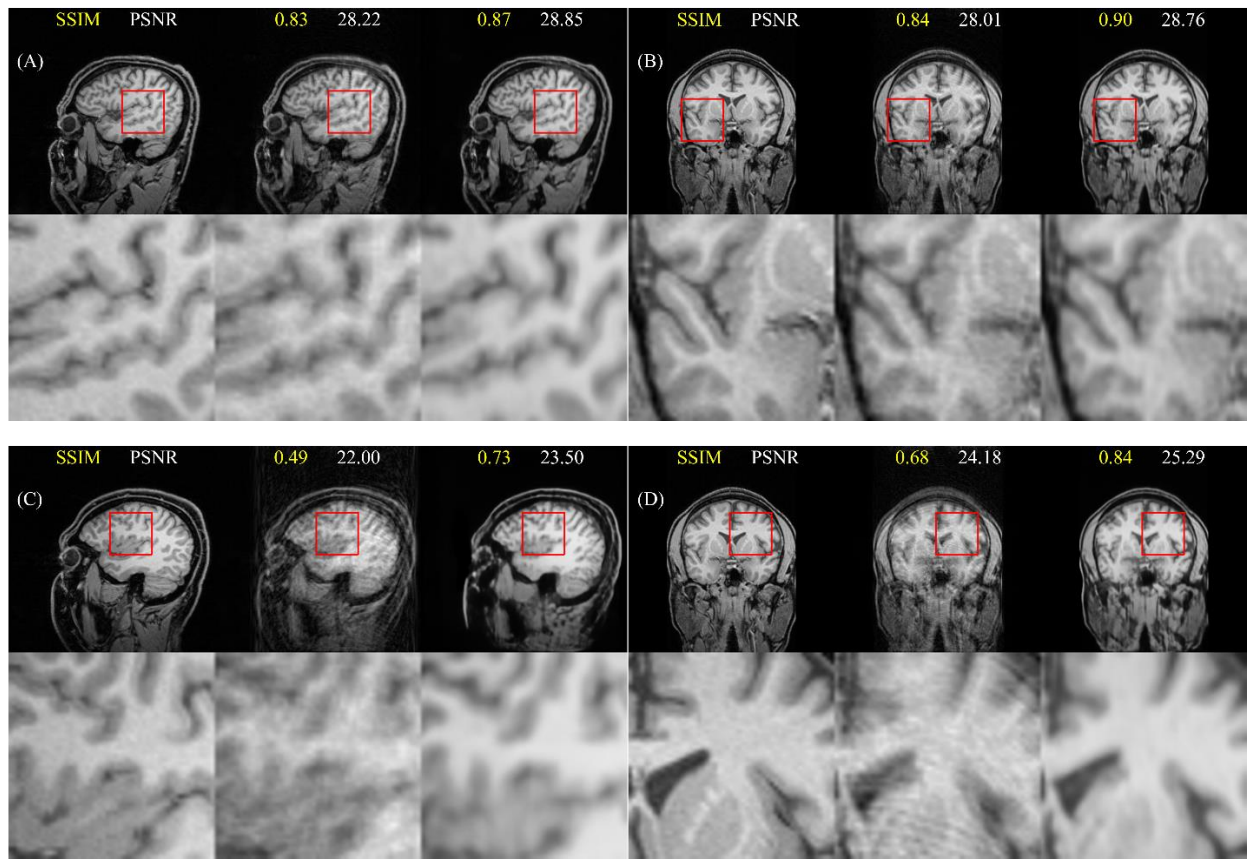

**Figure 7**. Results of cross-dataset generalization with motion-corrupted MP-RAGE images of saggital (A, C) and coronal orientations (B, D). In each subfigure, the first row shows the motion free image, motion-corrupted image, and images corrected by the MC-Net. The second row shows zoom in of the region of interest (ROI) within the red rectangle. Yellow and white numbers represent the SSIM and PSNR relative to the motion-free image.

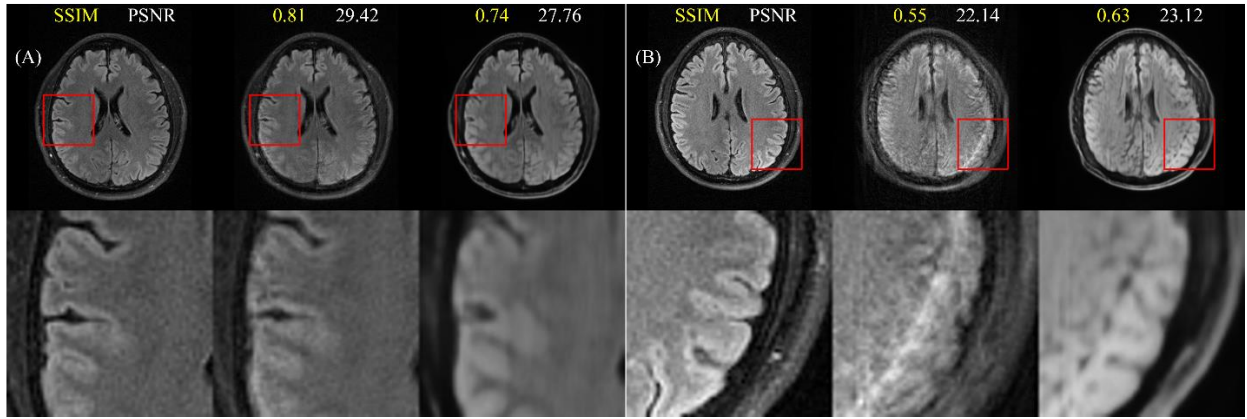

**Figure 8**. Results from two T2-weighted (FLAIR) images using simulated motion. The left set (A) was corrupted with relatively minor motion and the right set (B) with more severe motion. Note the appearance of false anatomical "features" (yellow arrows). Within each set, images in each column are original images, corrupted image, and outputs from MC-Net (left to right).

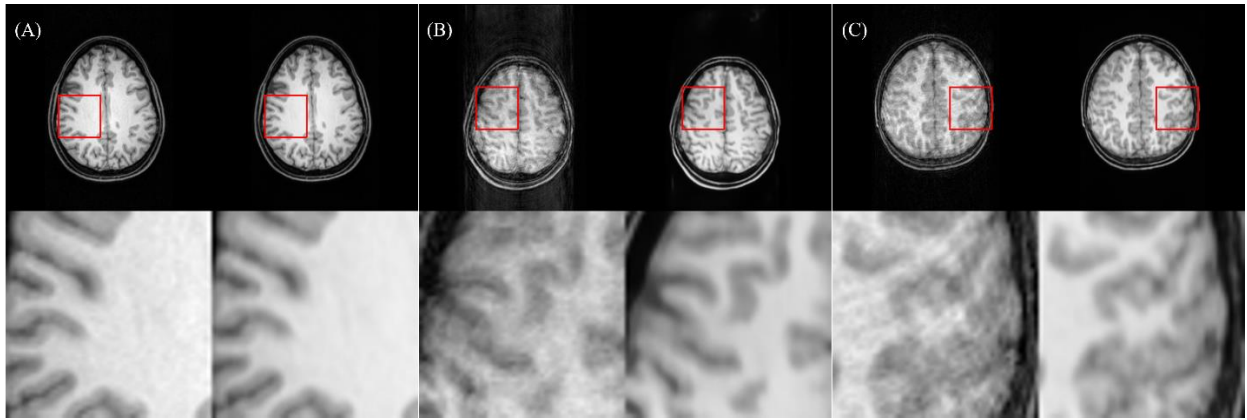

**Figure 9**. Examples of images with real motion (non-simulated) artifacts. From left to right (A to C), the severity of motion artifacts is increasing. Each set shows the original motion-corrupted image (left) and outputs from MC-Net (right).

SUPPLMENTARY

|      | Model | Clean Image | L1 | L1 + TV | Two-stage |
|------|-------|-------------|-----|---------|-----------|
| SSIM | U     | 1           | 0.959±0.011 | 0.961±0.009 | ***0.967±0.008*** |
| PSNR | U     | Inf         | 36.697±1.216 | 36.445±1.080 | ***37.403±1.168*** |
| SSIM | U+O   | 1           | 0.999±0.000 | *0.999 ± 0.001* | *0.999 ± 0.001* |
| PSNR | U+O   | Inf         | 47.004±2.015 | ***47.637 ± 2.713*** | 45.490 ± 1.833 |

TABLE 2. SSIM (mean ± std) and PSNR (mean ± std) of the motion-free images, and corrected images processed using L1, L1 + TV, and the two-stage multi-loss function. L1 denotes models trained with L1 loss, L1 + TV denotes models trained with L1 plus TV losses at one stage, and "Two-stage" denotes models trained with L1 loss at first stage and with L1 plus TV losses at second stage. U and U+O represent UNet-like structure and UNet-like structure with input-to-output concatenation [2].